\documentclass[preprint,12pt]{elsarticle}

\usepackage{amssymb}
\usepackage{graphicx}
\usepackage{xcolor}
\usepackage{textcomp}

\begin{document}

\begin{frontmatter}



\title{An upgraded focal plane detector for the MAGNEX spectrometer}


\author[1]{D. Torresi\corref{cor1}}
\ead{domenico.torresi@lns.infn.it}
\author[1]{O. Sgouros}
\author[1]{V. Soukeras}
\author[1]{M. Cavallaro}
\author[1,2]{F. Cappuzzello}
\author[1]{D. Carbone}  
\author[1]{C. Agodi}
\author[1,2,3]{G. A. Brischetto}
\author[1,2]{S. Calabrese}
\author[1,2]{I. Ciraldo}
\author[1]{N. Deshmukh\fnref{fn1}}
\author[1,4]{A. Hacisalihoglu}
\author[1,2]{L. La Fauci}
\author[1,2]{A. Spatafora}
\author{for the NUMEN collaboration}

\address[1]{INFN - Laboratori Nazionali del Sud, Catania, Italy}
\address[2]{Dipartimento di fisica e astronomia - Universit\'a di Catania, Catania, Italy}
\address[3]{Centro Siciliano di Fisica Nucleare e Struttura della Materia (CSFNSM), Catania, Italy}
\address[4]{Akdeniz University, Antalya, Turkey}

\cortext[cor1]{Corresponding author}
\fntext[fn1]{present address: Nuclear Physics Division, Saha Institute of Nuclear Physics, Kolkata, 
West Bengal 700064, India}

\begin{abstract}

An upgraded and improved version of the focal plane detector (FPD) of the large-acceptance magnetic
spectrometer MAGNEX is described here. The FPD consists of a tracker operating at low pressure and 
of a silicon detector wall. Thanks to a different geometry of the electron multiplication and 
induction elements, the new detector guarantees a superior signal to noise ratio, resulting in a 
more accurate tracking and a cleaner identification of the detected heavy ions. The new detector 
has been tested by using a $^{18}$O beam at an energy of 84 MeV. A description of the new FPD that 
pinpoints the innovation and the obtained performances is given and discussed in details.
\end{abstract}

\begin{keyword}
Magnetic spectrometer \sep Focal plane detector \sep MAGNEX

\end{keyword}

\end{frontmatter}

\section{Introduction}
   \label{intro}
   Modern nuclear physics studies often require to join the advantages of the magnetic spectrometry 
   (strong rejection factors of reaction products, zero-degree measurements, etc.) with the 
   possibility to measure heavy ions. Moreover, in order to explore rare nuclear processes 
   \cite{PhysRevLett.116.052501,CappuzzelloEPJA2018,PEREIRA2012426}, modern magnetic spectrometers 
   have grown toward large acceptance in momentum and solid angle as well as large dynamic range in 
   mass and energy \cite{KOBAYASHI2013294,UESAKA20084218,STEFANINI2002217,REJMUND2011184}. An 
   example of this class of devices is the MAGNEX spectrometer 
   \cite{CappuzzelloEPJA2016,CAVALLARO2020334,Cappuzzello2011book} installed at the INFN-LNS 
   laboratory in Catania. Its magnetic structure, based on a vertically focusing quadrupole and a 
   horizontally dispersing and focusing bending magnet, ensures a momentum acceptance of about 
   24\% and a solid angle of 50 msr. The development of large-acceptance magnetic spectrometers 
   requires more advanced Focal Plane Detectors (FPD). These have to provide not only an 
   unambiguous particle identification, but also an accurate three-dimensional tracking of the ions 
   trajectory downstream of the magnetic elements
   \cite{Lazzaro2005,CAPPUZZELLO201174,PULLANHIOTAN2008343,Montanari2011}.
   The FPDs are, therefore, the heart of the modern large-acceptance magnetic spectrometers.
   In the present paper the upgrade of MAGNEX FPD \cite{Cavallaro2012} will be described.
   The new FPD has been developed with the aim to improve the tracking and the identification
   performances.
   
   The paper is organized as follows: Section 2 describes  the design of the new detector, 
   emphasizing the new adopted solutions compared to the previous ones; Section 3 presents a 
   detailed characterization of the new FPD response to beam generated ions; Section 4 summarizes 
   the results and our conclusions. 

\section{Focal plane detector design}
   \label{sec:1}
   The MAGNEX FPD consists of two sections: a gas tracker sensitive to the energy loss of
   the ions and a    stopping wall for the measurements of their residual energy \cite{Cavallaro2012}. 
   The gas tracker is a 
   proportional drift chamber with a total active volume of 1360$\times$200$\times$90 mm$^3$, 
   divided in six sections that are six independent position-sensitive proportional counters, whereas 
   the stopping wall, placed behind the gas tracker, is made of 57 silicon pad detectors 
   covering an area of 1360$\times$200~mm$^2$, see Fig. \ref{fig:lateralFPD}.
    
   \subsection{The gas tracker}
     The gas tracker is 
     contained in a vacuum chamber that is isolated from the high-vacuum upstream region by a large 
     mylar\textsuperscript{\textregistered} 
      window (220$\times$920 mm$^2$) with typical thickness 
     ranging from 1.5 to 6 $\mu$m, depending on the pressure filling the chamber.
     The active region is filled with 99.95\% isobutane at pressure that ranges from few mbar up 
     to several tenths of mbar. The use of pure isobutane guarantees a fast drift velocity and 
     very good operational stability, 
     \cite{SCHMIDT1986579,SHARMA201298,Leo}. 
     In order to avoid further dead layer there is no exit window, the silicon detectors 
     of the stopping wall are therefore embedded in the gas. The FPD 
     vacuum chamber is movable of $\pm$0.08 m along the optical axis of the spectrometer to allow 
     to translate the focal plane according to different focus conditions.
     In order to reduce the effect of the chromatic aberrations the FPD is tilted at an angle
     $\theta_{tilt}$=59.2$^{\circ}$ relative to the optical axis direction \cite{CUNSOLO200256}.
 
\begin{figure}
      \centering
      \includegraphics[width=0.8\textwidth]{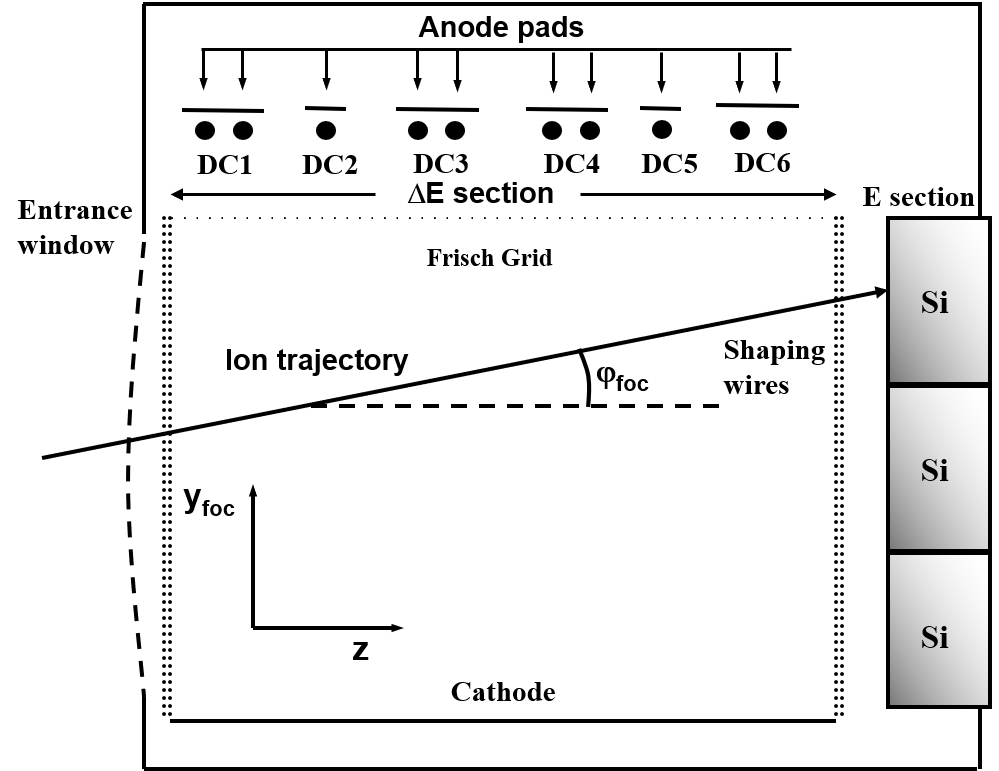}
      \caption{Sketch of the lateral view of the new MAGNEX FPD.}     
      \label{fig:lateralFPD}%
     \end{figure}

     In the tracker, sketched in Fig. \ref{fig:lateralFPD} it is possible to identify three 
     different regions. A drift region defined by the cathode and the Frish grid, a 
     multiplication region which extends between the Frish grid and the proportional wires,
     and an induction region which extends between the DC wires and the segmented anode.
          
     The drift region is defined by the cathode, an aluminum plate 1200$\times$90~mm$^2$ 
     large that is usually biased at voltage values ranging between -900 and -1500 V and a Frish grid,
     that is made of ten gold-plated tungsten wires with a diameter of 50 $\mu$m placed at a 
     distance of 5 mm one from each other. In order to make as uniform as possible the electric 
     field of the drift region and to shield the inner electric field from the high voltage 
     applied to the silicon detectors at the stopping wall \cite{CavallaroPHD}, the active area of 
     the tracker is surrounded by a double partition grid consisting of 41 couples of rings made of 
     gold-plated wires.
           
\begin{figure}
      \centering
      \includegraphics[width=0.8\textwidth]{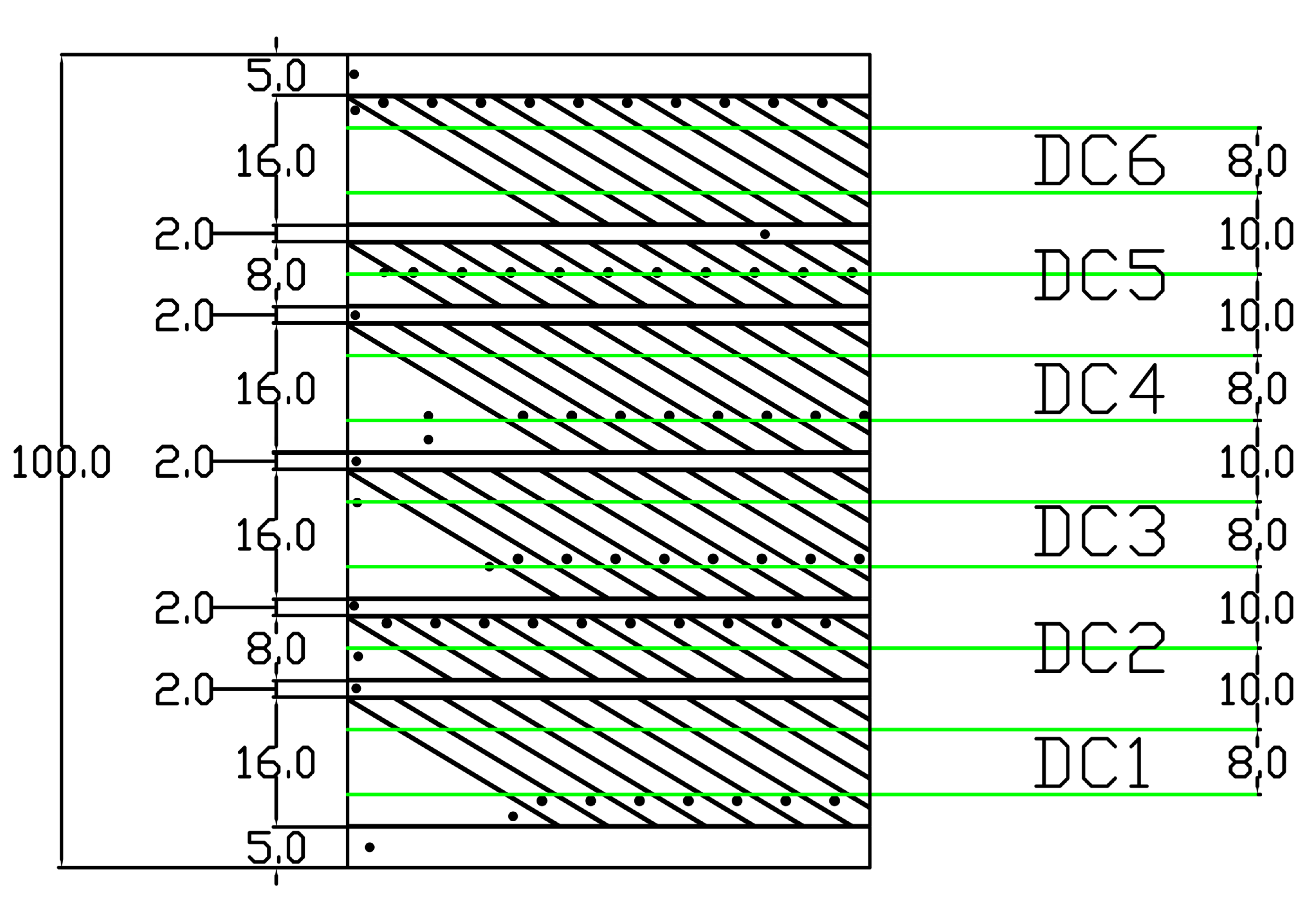}
      \caption{Scale drawing of a small portion of the segmented anode (bottom view). In the drawing 
      the six strips of the anode that corresponds to the six drift chambers DCs are shown . 
      Each strip is further segmented in pads that are tilted of $\theta_{tilt}$=59.2$^{\circ}$.    
        In green the proportional wires are shown.     }

      \label{fig:PADview}%
     \end{figure}

     The multiplication region is 20 mm high and is defined by the Frish grid and the plane 
     where 10 proportional wires are located. Each of the proportional wires is 50~$\mu$m 
     thick and is made of gold-plated tungsten. Such wires are biased to a voltage usually 
     in the range between +500~V and +1300~V, provided by a common power supply. The ten wires are 
     shared among the six proportional counters DC$_i$,~i=1,$\ldots$~6. DC$_2$ and DC$_5$ have 
     just a single wire while the the other DCs have two proportional wires as shown in Fig.
     \ref{fig:PADview}. An additional partition grid similar to the one used in the drift region 
     is present with the aim to reduce the border effects of the electric field in the 
     multiplication region.
          
     The induction region is defined by the plane where the 10 proportional wires lay and the 
     anode: the latter consisting of a segmented read-out plane. In Fig. \ref{fig:PADview} a bottom 
     view of the anode is
     shown. The anode is divided in 6 longitudinal strips, one for each DC, being the strip 
     corresponding to DC$_2$ and DC$_5$ 8 mm wide while the others are 16 mm wide. Each strip is 
     further segmented in pads (221 for DC$_2$ and DC$_5$ and 223 for DC$_{1,3,4,6}$) oriented along the 
     spectrometer optical axis, that is with an angle equal to $\theta_{tilt}$, see Fig. 
     \ref{fig:PADview}.

   \subsection{The silicon stopping wall}
     The silicon stopping wall is embedded in the gas filling the tracker to avoid further 
     dead layers. 
     The stopping wall is made of 57 silicon detectors arranged in 19 columns. Each detector is  
     50$\times$70 mm$^2$ wide and 500~$\mu$m or 1000~$\mu$m thick depending on the range of the 
     ions to stop. They are mounted orthogonally to the optical axis of MAGNEX in order to minimize 
     the effective dead layer.
     The closest distance of the silicon detector from the active area of the tracker is 15 mm
     that is enough to avoid interference with the electric field of the drift region of the tracker.\\

     \subsection{The working principle}
     When an incident particle crosses the entrance 
     window entering the drift region, it generates a track of ions and electrons in the gas.
     The presence of a uniform electric field of about 
     50~V/cm in the drift region makes the ions drift towards the cathode and the electrons towards 
     the Frish grid, these last with a velocity of about 5 cm/$\mu$s \cite{SCHMIDT1986579}.
     After the Frish grid the primary electrons enter the multiplication region, are then accelerated 
     by the strong electric field generated by the proportional wires and the multiplication occurs. 
     Since the gas counters work in a proportional regime, the avalanches produced close to the wires
     generate a signal on the wires themselves which is proportional to the energy loss of the ion in 
     the gas. Therefore six independent measurements of the energy loss are available, one for each DC.

     In addition to the signal produced on the DC wires, the electron avalanche produced close 
     to the wires induces a charge in a given number of pads of the strip laying just above the wires. 
     The center of gravity of the charge distribution of the pads corresponding to a given DC 
     is extracted. The six centers of gravity are converted in horizontal position
     providing six independent measurements X$_i$ with $i=1,\ldots 6$. From them, the position 
     of the crossing point between the ion track and the focal plane X$_f$ as well as the horizontal 
     angle of the track $\theta_f$ is obtained.
     After crossing the gas tracker the ions hit the silicon detector stopping wall. 

     The vertical position is extracted by measuring the arrival time of the electron avalanches in 
     the wires taking advantage of the fact that the tracker works in a regime where the drift 
     velocity is almost constant in the whole volume of the detector.         
     Six vertical positions are extracted measuring the drift time of the primary electrons along
     in the drift region. The start signal is generated when the ion producing the track hits one
     of the silicon detector of the stopping wall. 
     The six vertical positions Y$_i$ with $i=1,\ldots 6$ are used to obtain the vertical
     position Y$_f$ on the focal plane detector and the angle $\phi_f$ of the ion track.     
     The 57 silicon detectors of the stopping wall, in addition to the start signal
     for the drift time measurements, provide also the residual energy of the ions that 
     is used mainly for identification purposes.
     
     Thanks to the very small dead layer, almost entirely due to the mylar\textsuperscript{\textregistered}
      window, the energy 
     threshold for the detection of charged particles crossing the FPD is about 0.5 MeV/u.

\section{Detector performaces}
   \label{sec:2}  
   The FPD was tested with a $^{18}$O beam delivered by the tandem MP at INFN Laboratori Nazionali 
   del Sud in Catania, Italy, at an energy of 84~MeV impinging 
   on thin gold and carbon targets. The interaction between the beam particles and the target 
   took place in the center of the MAGNEX scattering chamber. The angle between the
   optical axis of the spectrometer and the beam direction is named $\theta_{opt}$. 
   The spectrometer worked in full-acceptance mode which means an angular range ($\theta_{opt}$ 
   -5.2$^{\circ}) < \theta_{lab} <( \theta_{opt}$ + 6.3$^{\circ}$) \cite{CAVALLARO201146,CAVALLARO201177}. 
   Two collimators were 
   installed upstream the target in order to limit the spot size and angular divergence of the beam
   at the target position to 1.78 mm $\times$ 1 mrad in the horizontal direction and 2.8 mm 
   $\times$ 1.2 mrad in the vertical one. In some of the runs a multi-hole collimator was placed 
   206~mm downstream the target in order to select ejectiles with well defined trajectories. The 
   multihole collimator has 65 circular holes with a diameter of 1 mm, arranged in five rows 11.5 
   mm spaced and 13 columns 3 mm spaced. The ions coming from the target which enter in the 
   spectrometer acceptance were momentum analyzed and focused on the FPD. The FPD was filled with
   isobutane C$_4$H$_{10}$ at a pressure of 15 mbar. In the following section we will analyze the 
   main characteristics of the FPD, that is the resolution of the energy lost in the gas; the 
   angular resolution; the position resolution, and the particle identification performances.
      
   \subsection{Energy loss measurements}
   
   Even if the energy loss measurement of the new FPD is based on the  same principles as the old one
   Ref. \cite{Cavallaro2012}, some difference is present.
   In fact the old FPD was composed of five drift chambers, but, usually, only the central one, 
   the largest,    was used to measure the energy loss $\Delta$E.
   In the present FPD, there are six independent drift chambers with comparable active volume
   (DC$_1$, DC$_2$, DC$_3$, DC$_4$, DC$_5$, DC$_6$).	
   The total active volume of the six drift chambers is larger than the volume of the 
   single drift chamber used in the old FPD for the $\Delta$E measurement. This fact ensures
   higher charge 
   collection and therefore a better energy resolution for the new FPD, in the same ionization 
   conditions.

   Before making any further consideration on the $\Delta$E measurements we have to underline that 
   the FPD of a large-acceptance spectrometer should be designed to measure trajectories with very 
   different incident angles $\theta_f$ and therefore with different effective lengths inside 
   the FPD itself. 
   In the case of MAGNEX the FPD is tilted of $\theta_{tilt}$=59.2$^{\circ}$ relative to the central 
   trajectory, and $\theta_f$ (angle in the dispersive direction) ranges between a minimum value of 
   40$^{\circ}$ and a maximum value of 72$^{\circ}$. 
 
   Therefore the linear length of the trajectory inside the active volume of the FPD can range
   from about -35\% up to about +65\% of the central trajectory.
   The contribution coming from the non-dispersive direction is almost negligible since the 
   angle in $\phi$ ranges from -2$^{\circ}$ to +2$^{\circ}$ \cite{CAPPUZZELLO201174}.

   As a consequence the $\Delta$E 
   measurement is corrected for the effective thickness and normalized to the  
   energy loss of the reference trajectory angle $\theta_{tilt}$=59.2$^{\circ}$. Therefore we 
   introduce the quantity $\Delta$E$_{corr}$:
   \begin{equation}
      \Delta E_{corr}=\Delta E\frac{cos\theta_f}{cos\theta_{tilt}}
      \label{eq:1} 
   \end{equation}
   \begin{figure}
     \centering
     \includegraphics[width=1\textwidth]{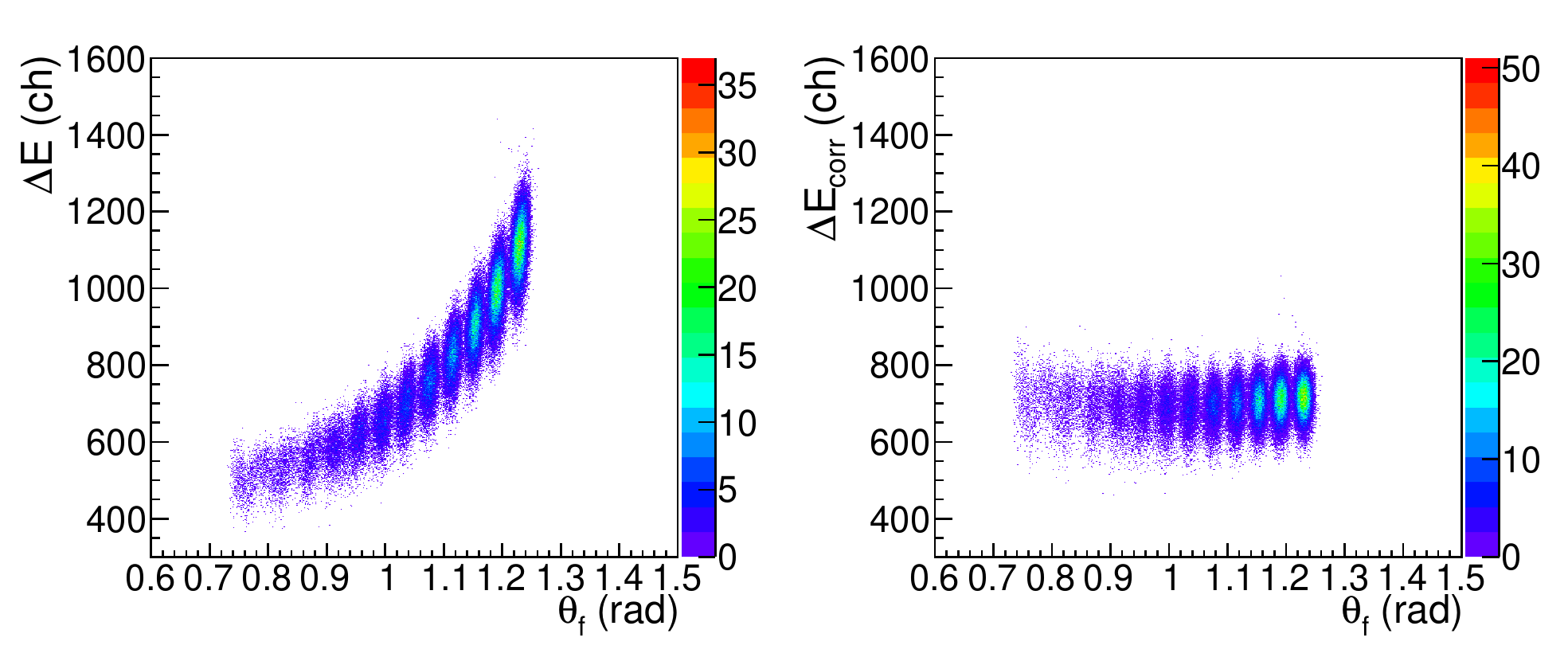} 
     \caption{Left pad Energy loss ($\Delta$ E$_{corr}$) as a function of $\theta_f$, right pad 
     energy loss ($\Delta$E$_{corr}$ corrected for the effective path as a function of $\theta_f$. 
     }     

     \label{fig:energyCorr}%
   \end{figure}

   To evaluate the effectiveness of this correction procedure 
   the elastic scattering of $^{18}$O beam at 84 MeV  
   on a 122 $\mu$g/cm$^2$ thick gold target was used. The FPD was working with 
   99.95\%-pure isobutane gas at a pressure of 15 mbar. The voltage applied to the cathode and wires were 
    -1000 V and +720 V respectively. The effect of such a correction is shown in Fig. 
   \ref{fig:energyCorr}. In the left panel $\Delta$E as a function of $\theta_f$ is shown and it is 
   evident the dependence of the measured energy loss from the horizontal incident angle $\theta_f$. 
   In right panel, where $\Delta$E$_{corr}$ as a function of $\theta_f$ is shown, the dependence of the 
   energy on the angle $\theta_f$ is removed as expected.
   It is, therefore, evident that a precise measurement of the angle $\theta_f$ is not only important 
   for reconstruction purposes but it is also critical for identification purposes using the $\Delta$E-E 
   technique.

\begin{figure}
     \centering
     \includegraphics[width=0.9\textwidth]{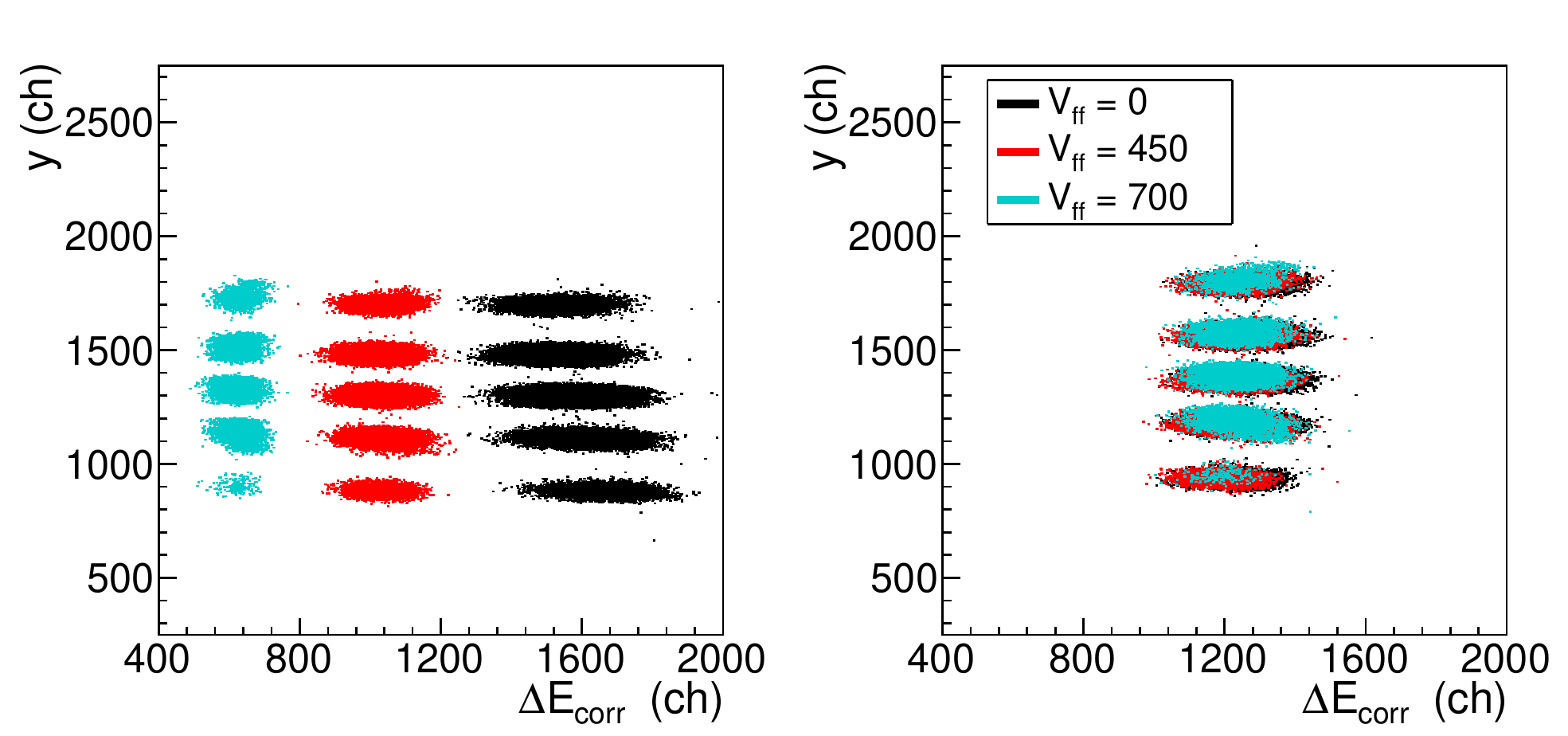}
     \caption{y-coordinate versus $\Delta$E lost in the gas for DC$_1$ on the left and for DC$_3$ on 
     the right. Each 
     value of the voltage applied to the fringing filed is represented as a different colour.}
     \label{fig:energyvsY}       %
   \end{figure}

   Border effects for the charge-collection efficiency are known to be present in particular 
   at the entrance and exit of the FPD, where the electric field could be not uniform.
   Such non uniformity can worsen the tracking performances and generate a dependence of
   the collected charge from the y-coordinate. 
   In order to mitigate such effects the voltage of the partition grid in the electron
   multiplication region was varied looking for the achievable best condition.
   The change of the energy channel for different values of the partition grid potential,
   all other parameters being equal, is shown in Fig. \ref{fig:energyvsY} where $y$ vs 
   $\Delta$E$_{corr}$ is 
   plotted for DC$_1$ (example of a border DC) and DC$_3$ (example of a central DC).
   Different colors correspond to different bias applied to partition grid potential. 
   For the case of the central 
   DC$_3$ there is no effect, for DC$_1$ there is a strong effect that reduces the 
   amplitude of the energy loss signals and virtually removes the dependence on the y coordinate.

\begin{figure}
     \centering
     \includegraphics[width=0.9\textwidth]{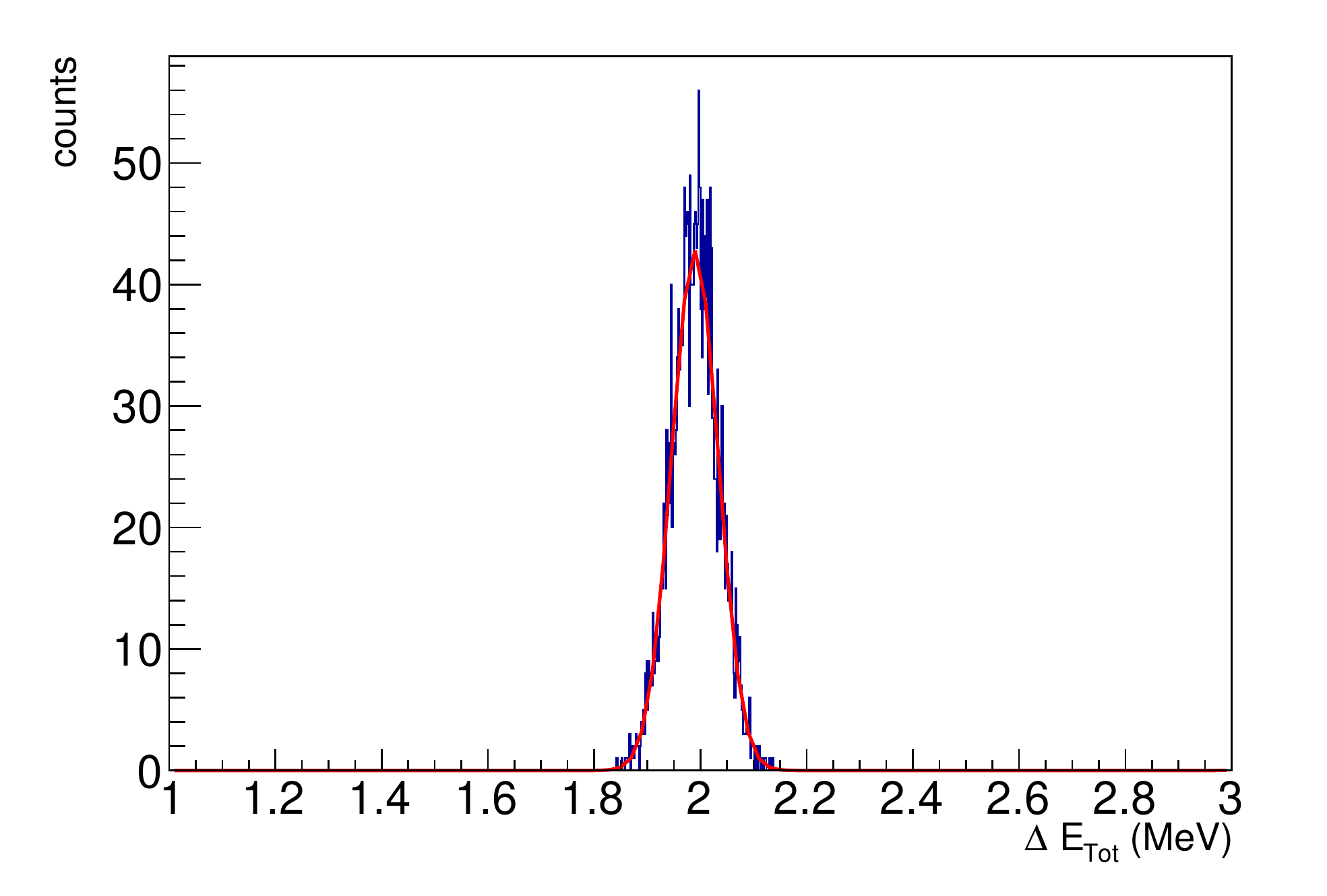}
     \caption{Example of distribution of the total energy lost in the six DCs using a multi-hole 
     collimator. The FWHM extracted is about 5\%. See text for more details.}
     \label{fig:energy_distr}       %
   \end{figure}

   In order to study the intrinsic resolution of the energy loss ($\Delta E$) measurement, the 
   elastic scattering data collected with the multiple hole collimator were analyzed. Moreover, 
   the events corresponding to each hole of the collimator were selected with a condition in 
   $\theta_f$ ($\Delta\theta_f  <$ 2 mrad) and y ($\Delta y <$ 2 mm) in order to remove the 
   dependence on the ion trajectory. While the FWHM of the $\Delta$E$_{corr}$ distribution for each single 
   DC$_i$ was about 10\%, the overall resolution obtained by summing all the energy measured by the 
   six DCs is about 5\%. The total energy here considered is given by 
   $$\Delta E_{tot}=\Delta E_{corr1}+\Delta E_{corr2}+\Delta E_{corr3}+\Delta E_{corr4}+\Delta E_{corr5} +\Delta E _{corr6}$$
   The result does not depends significantly on the particular value of $\theta$ and y. An example of 
   such calibrated energy spectrum is shown in Fig. \ref{fig:energy_distr}.

   \subsection{Particle identification}
   In this section the particle identification procedure and typical results for the new MAGNEX FPD 
   are shown.
   \begin{figure}
     \centering
     \includegraphics[width=0.9\textwidth]{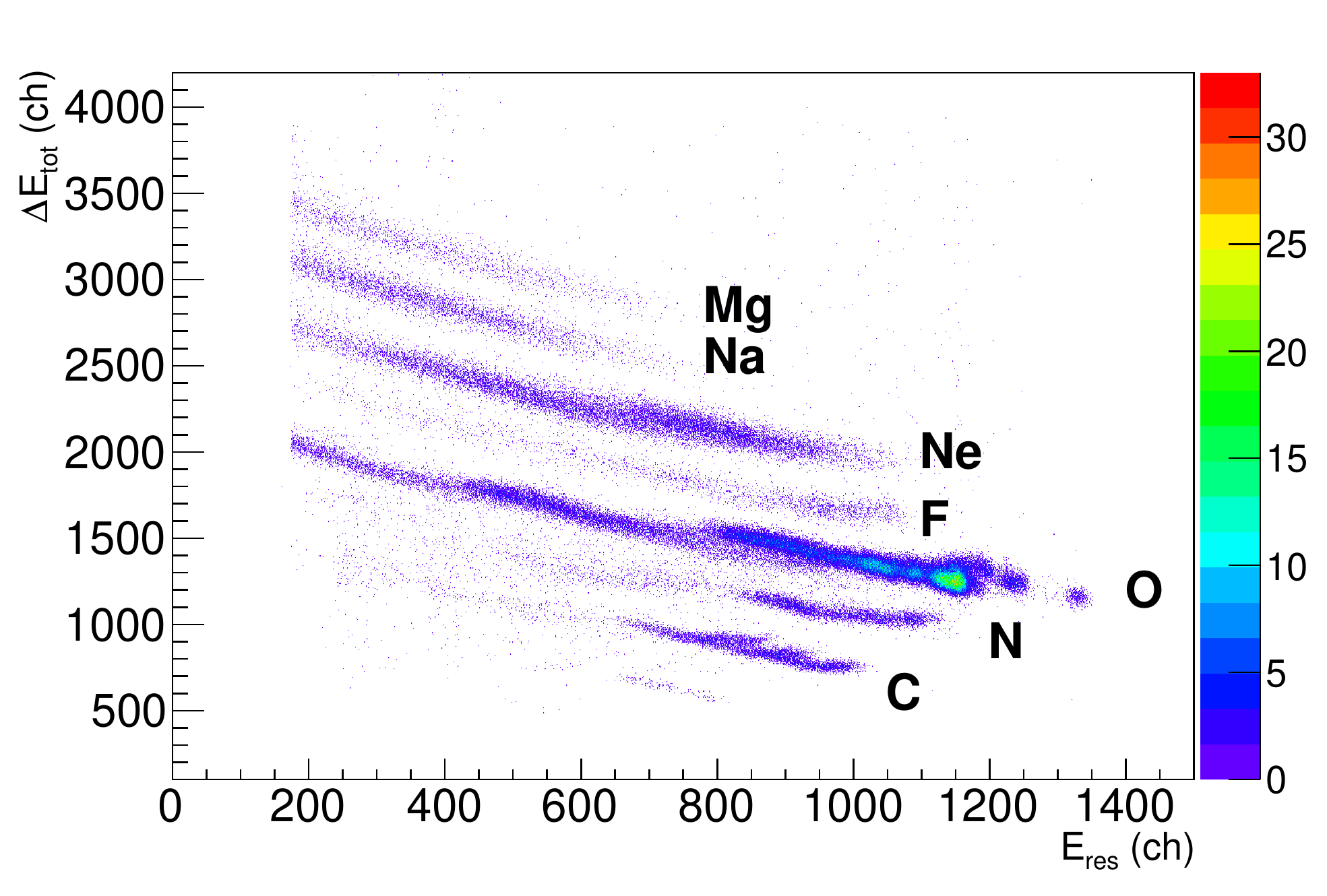}
     \caption{Typical $\Delta E_{tot}$ - E$_{res}$ plot for single silicon detector. 
     Loci of isotopes from 
     carbon to magnesium are well separated.}
     \label{fig:deltaEE}       %
   \end{figure}
   The data were taken using a $^{18}$O beam at 84 MeV impinging on a 238 $\mu$g/cm$^2$ thick
   $^{12}$C target. The spectrometer was set at $\theta_{opt}$=10$^{\circ}$. 
   
   In Fig. \ref{fig:deltaEE} the energy lost in the drift chamber corrected for the incident angle
   ($\Delta E_{corr}$) versus the residual energy E$_{res}$ measured by one of the silicon detectors 
   is shown. The different loci correspond to different values of the atomic number Z. 
   \begin{figure}
      \centering
      \includegraphics[width=0.9\textwidth]{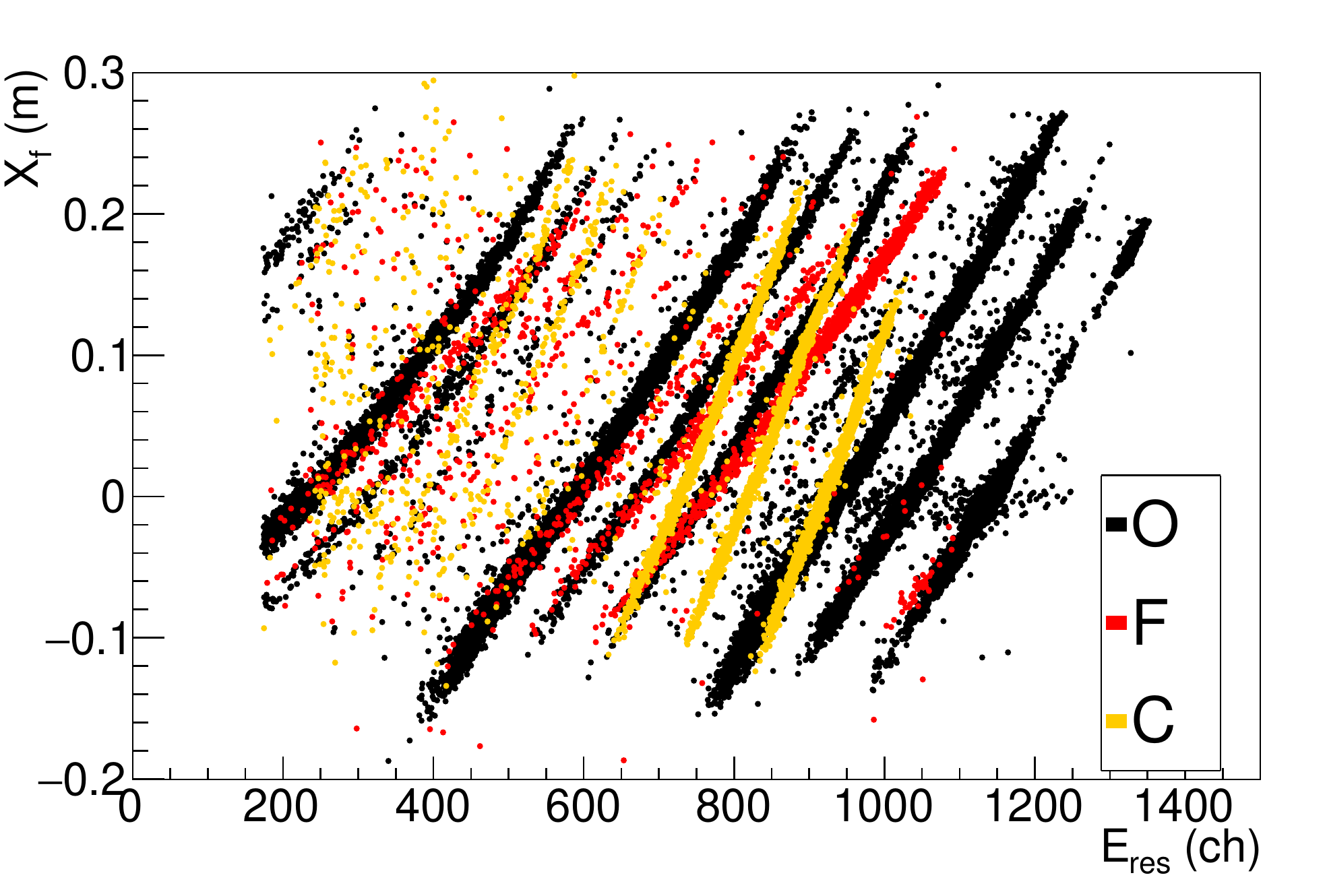} 
     \caption{Typical X$_{f}$ vs E$_{res}$ plot for a single silicon detector. 
     }
     \label{fig:xfitEres}       %
   \end{figure}
   
   In Fig. \ref{fig:xfitEres} the position on the FPD X$_{f}$ versus the 
   residual energy E$_{res}$ in one of the silicon detectors is shown.
   In this plot ions with the same Z gather on lines with the same slope, larger (smaller) slopes 
   correspond to larger (smaller) Z. Ions having same Z but different masses can be identified as
   lines with the same slope but shifted at smaller (larger) X$_{f}$ value for low (high) masses.
   This is the result of the relationship between the kinetic energy (related to the parameter E$_{res}$) and 
   the magnetic rigidity (related to the parameter X$_f$) of the charged particles traversing a 
   magnetic spectrometer. This identification technique has been reported in details in Ref.
   \cite{CAPPUZZELLO2010419}.
   \begin{figure}
     \centering
     \includegraphics[width=0.9\textwidth]{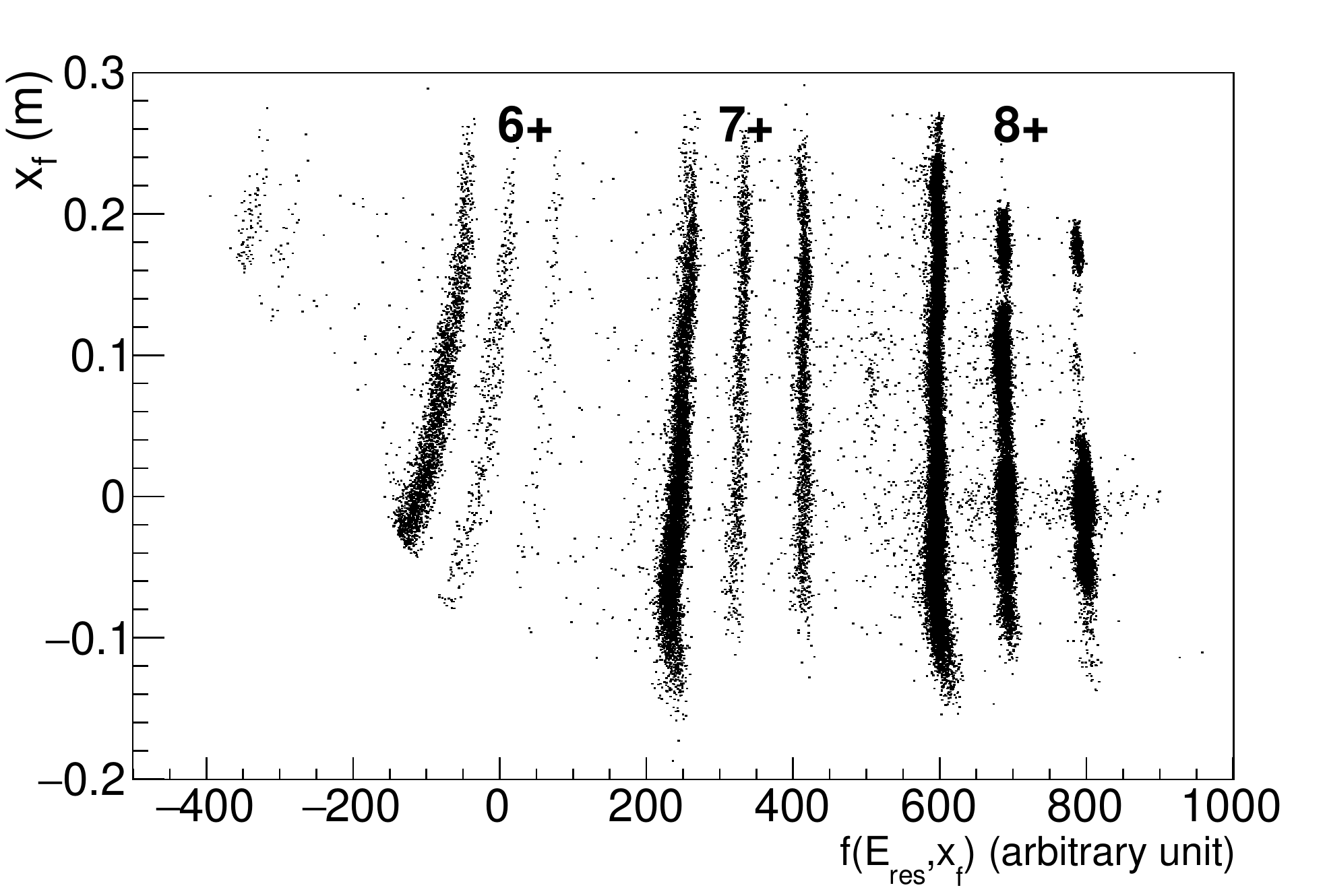}  
     \caption{X$_{f}$ vs a function of the residual energy and the position
     on the focal plane f(E$_{res}$, x$_{f}$)=$ E_{res}-(a\times x_{f}^2+b\times x_{f}+c)$ 
     for oxygen ions.  It is possible to identify ions with different charge states: 6$^+$, 7$^+$, 8$^+$.
     }
     \label{fig:rectified_a}       
   \end{figure}  
   \begin{figure}
    \centering
    \includegraphics[width=0.9\textwidth]{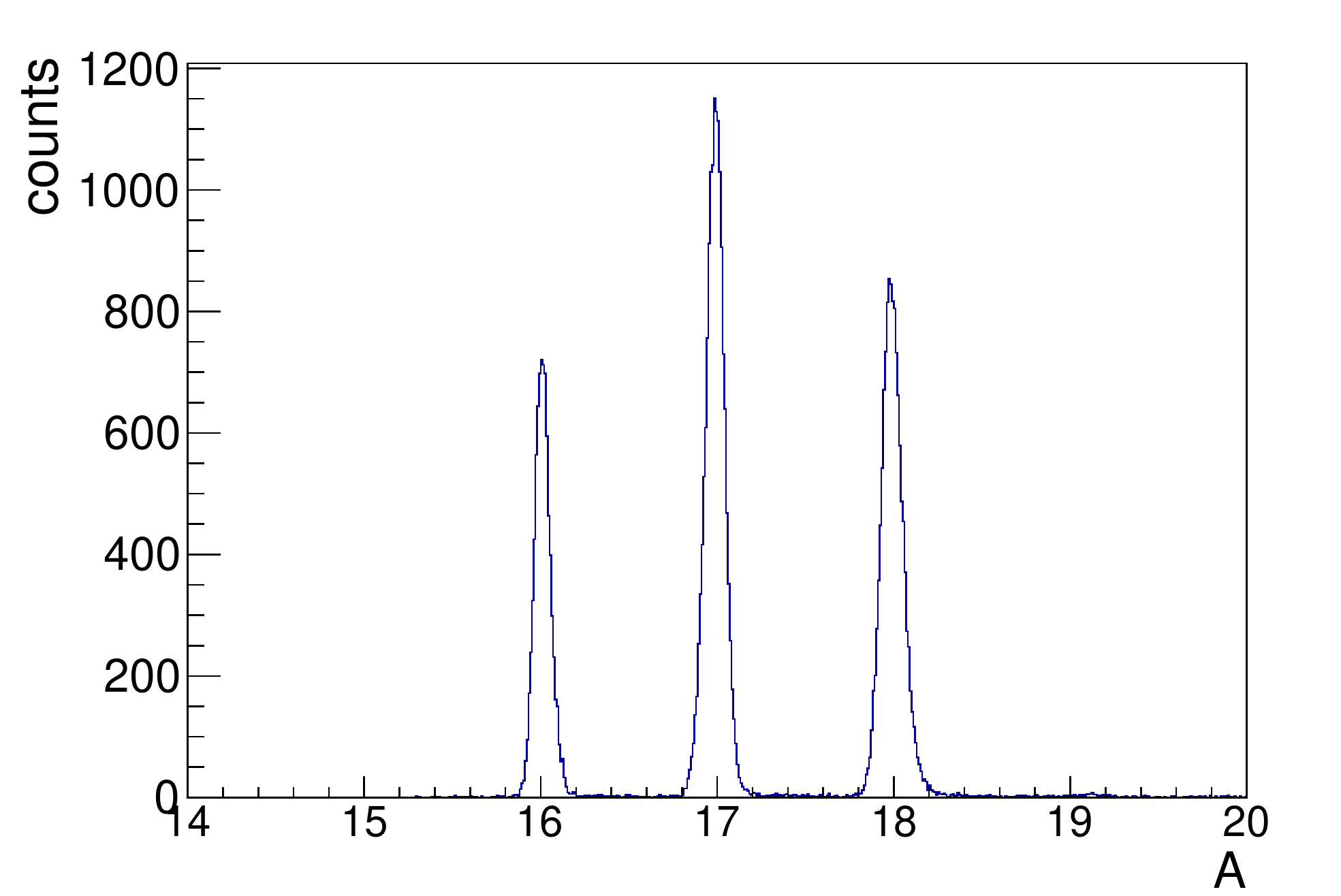}
     \caption{Mass distribution estracted from Fig. \ref{fig:rectified_a} for the oxygen ions: 
     A= 16, 17, 18 with a 8$^+$ charge state.}
     \label{fig:rectified_b}          
   \end{figure}
   These lines can be transformed in vertical lines using a rotation in the plane X$_{f}$ - 
   E$_{res}$ as shown in Fig \ref{fig:rectified_a}. After the rotation, the loci corresponding
   to oxygen isotopes with charge state 8$^+$ are selected, thus a cut on X$_f$ is performed and,
   eventually, they are projected on the x-axis (Fig. \ref{fig:rectified_b}). From this figure
   the mass resolving power of the FPD can be determined; it is defined as $R=M/\Delta M$, where M 
   is the mass 
   of the central peak and $\Delta$M is the FWHM of such peak. For example for oxygen ions the    
   resolving power, in this specific working conditions, is M=1/136, as shown in Fig.
   \ref{fig:rectified_b}.
   
   \subsection{Position and angle measurements in the dispersive direction}
     
   The position along the horizontal direction is determined by means of the distribution of charge
   induced on the pads. The number of hit pads is usually between 5 and 25 depending mainly on the 
   incident angle of the ion $\theta_f$ and its energy loss. A generalization of the center of 
   gravity method \cite{CHARPAK1979455} is used to extract the centroid of the charge distribution 
   as discussed in Refs. \cite{CARBONE2012,Cavallaro2012}.
   
   The position resolution has been estimated using the multi-hole collimator. Several effects 
   can influence the estimate of the position resolution like: the size of the collimator holes,
   the beam-spot size, multiple scattering in the detector itself, and straggling in the 
   entrance window. 
   These effects can be partially compensated by considering the difference between two
   coordinates as, for example,  X$_i$-X$_3$ instead that a single coordinate X$_i$. X$_3$ was 
   chosen as reference wire since, being a central drift chamber, it is less sensible to border 
   effects that could affect the edge drift chambers.
   
\begin{figure}
      \includegraphics[width=0.9\textwidth]{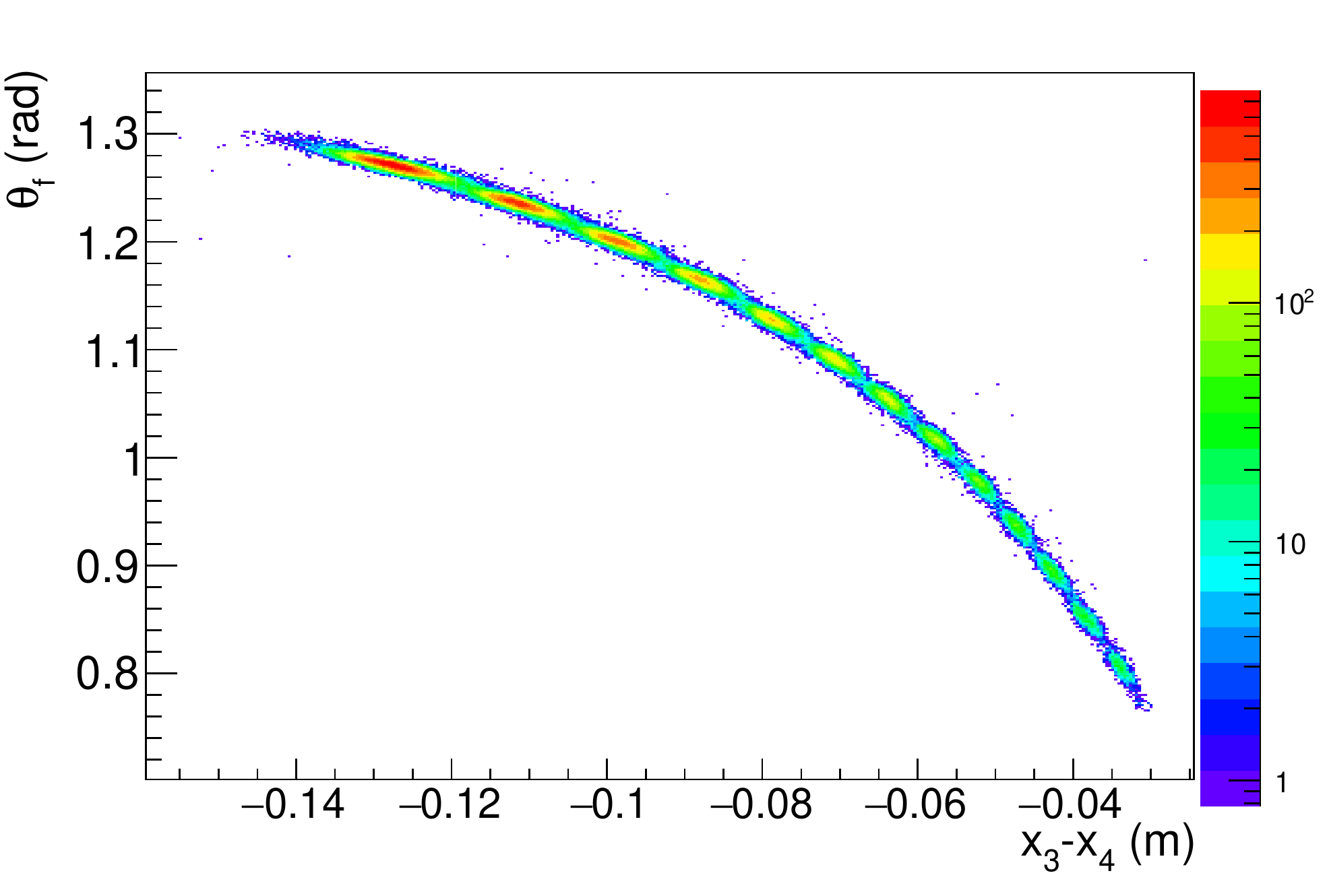}
     \caption{$\theta_f$ angle versus the $\Delta$x=X$_3$-X$_4$.}
     \label{fig:DeltaXvsTheta}       %
   \end{figure}

   In Fig. \ref{fig:DeltaXvsTheta} the angle $\theta_f$ versus X$_3$-X$_4$  is shown after
   a gate on the elastic scattering was applied.
   Each spot of the plot corresponds to a bunch of trajectories passing through the same hole
   having a given average angle $\theta_f$ at the focal plane
   As $\theta_f$ increases the X$_3$-X$_4$ distribution of each spot becomes broader.
    
\begin{figure}
     \centering   
     \includegraphics[width=0.9\textwidth]{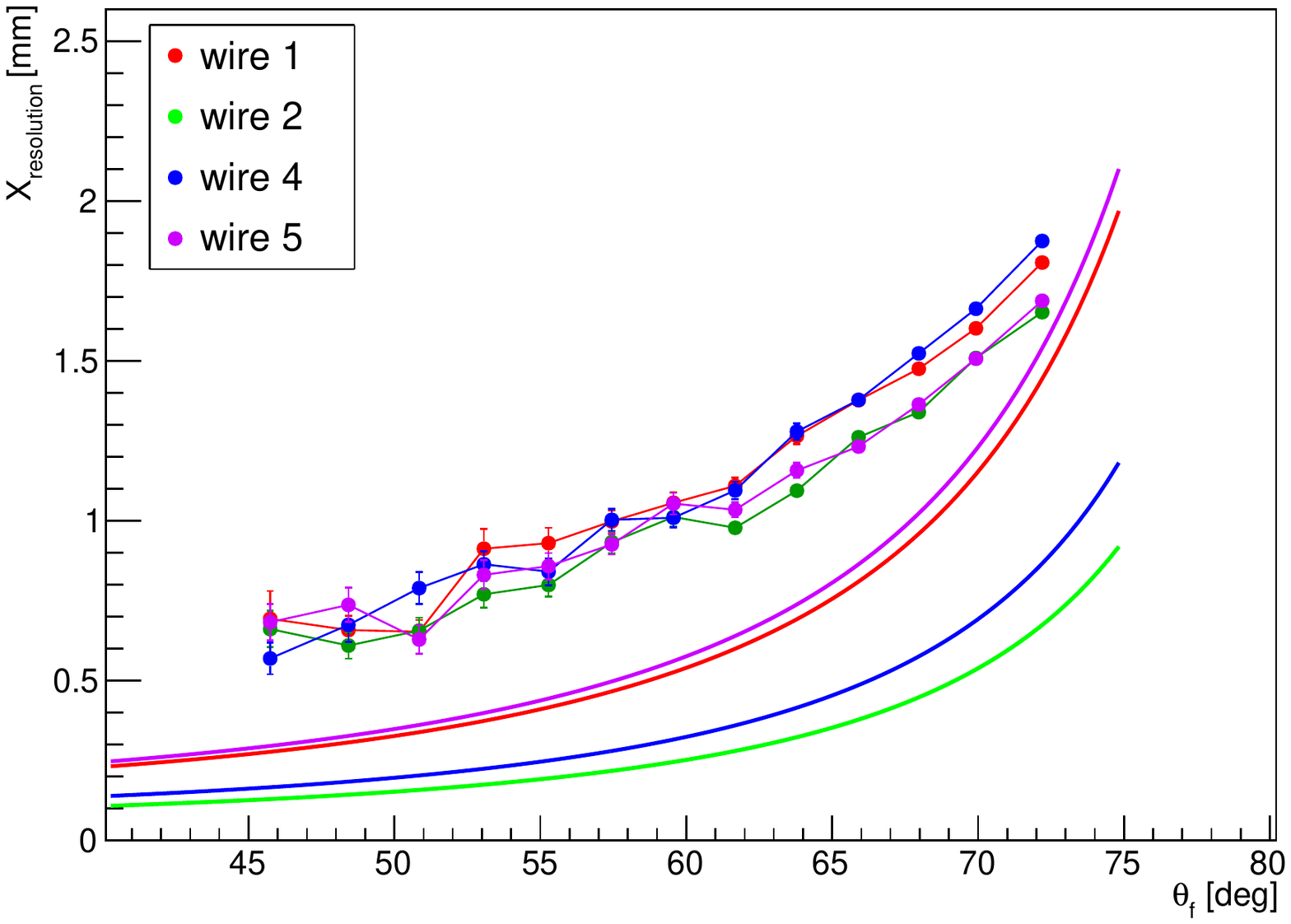}  
     \caption{Horizontal resolution (FWHM) as a function of the angle $\theta_f$. The lines represent the 
     geometrical contribution to the resolution due to the selected angle.}
     \label{fig:resolution}       
   \end{figure}

   For each spot of Fig. \ref{fig:DeltaXvsTheta} a narrow gate of 2 mrad in $\theta_f$ was applied
   and the position distributions of all the X$_i$ variables where obtained.
   Such angular spread corresponds, for example, to a position spread 
   of about 0.12~mm at $\theta$=40$^{\circ}$ and 0.65~mm at $\theta$=72$^{\circ}$. This geometrical
   contribution to the overall width of the position distribution is shown in Fig. 
   \ref{fig:resolution} as a continuous line for all the X$_i$, 
   while points correspond to the experimental FWHM measured at each angle. The horizontal
   resolution X$_i$-X$_3$ was estimated for all the wires. No significant difference 
   between different DC$_i$ was found.
   The final resolution is estimated to be around 0.6~mm for each wire.
         
   All the coordinates X$_i$, i=1,$\ldots$ 6 are used, together with the longitudinal coordinates
   Z$_i$, i=1,$\ldots$ 6 for the definition of the horizontal angle $\theta_f$. The use of six 
   points guarantees a better precision compared to the previous FPD that used just four points. 
   
\begin{figure}
     \centering
     \includegraphics[width=0.9\textwidth]{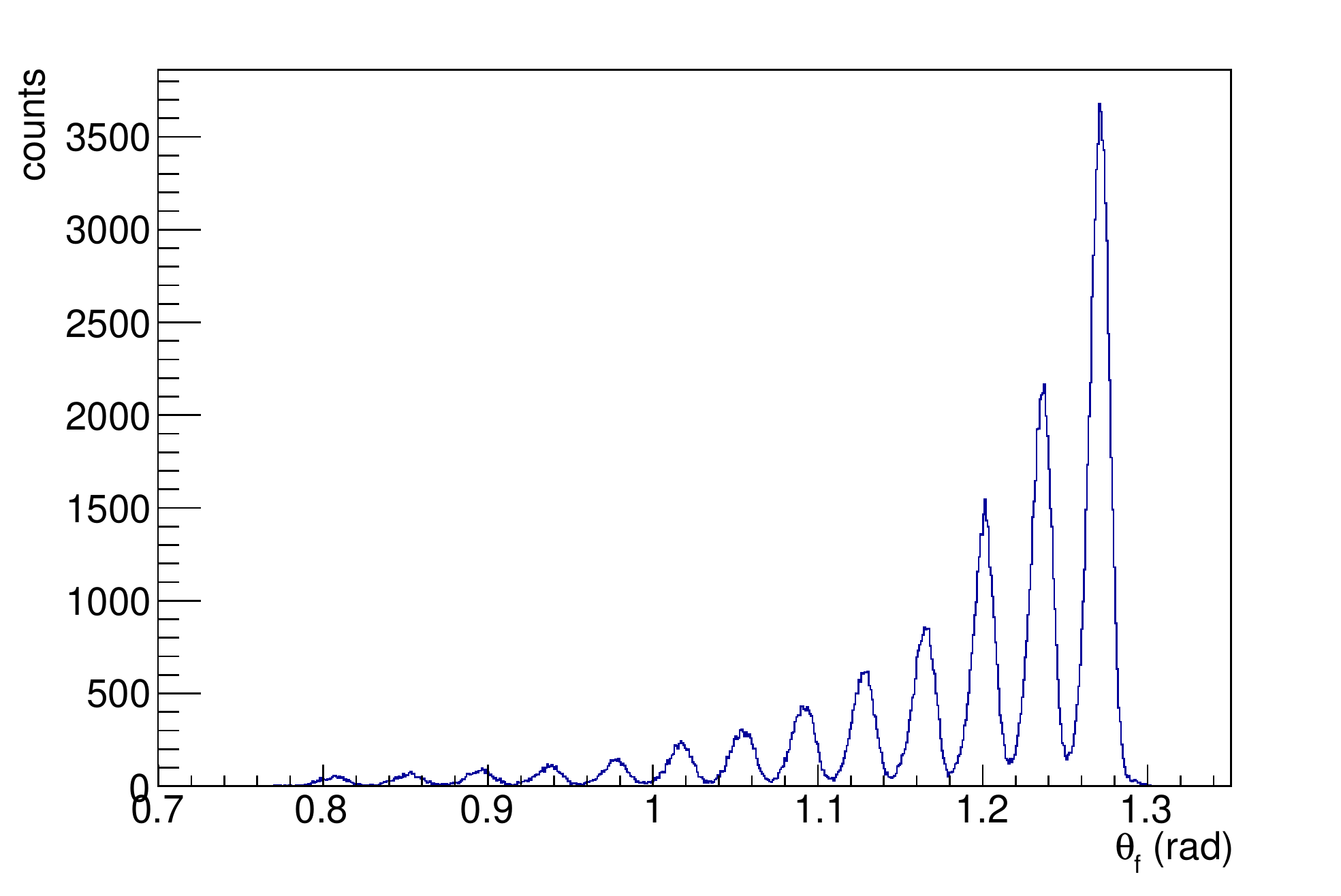}
     \caption{Plot of the $\theta_f$ angle after a selection in the elastically scattered
     $^{18}$O ions and in the trajectories passing through the middle-raw holes of
     the multi-hole collimator.}
     \label{fig:ThetaF}       
   \end{figure}

   The plot of measured  $\theta_f$ is shown in Fig. \ref{fig:ThetaF}. Each peak in the 
   plot corresponds to trajectories passing through one of middle-raw hole of the multi-hole 
   collimator. The difference in counts for each peak is due to different cross section of the 
   elastic scattering at different angles.

   The method used to estimate the angular resolution is described in detail in the following. The 
   first step consists in selecting a group of trajectories by applying a gate on 
   the first and the last DCs, (i.e. DC$_1$ and DC$_6$). 
   The trajectories are selected with an average $\theta_f$ value of 1000 mrad and a FWHM of 3.2 mrad.
      
   In the second step the trajectory of the events selected in the 
   first step are extracted, using the $X_i$ obtain from the other four DCs, thus excluding X$_1$ 
   and X$_6$. The angular distribution of such events has a spread of 3.9 mrad (FWHM), taking 
   into account the initial spread of
   selected trajectories it is possible to extract a value for the $\theta_f$ resolution of 
   the tracker of 2.2 mrad (FWHM). The resolution here extracted should be considered as an upper 
   limit of the actual $\theta_f$ resolution, since the resolution of the full tracker is done using 
   six X-positions for each events instead of the previous four X-positions we were using for the $\theta_f$ 
   resolution estimation. The resolution here obtained 
   is much better than the previous detector, where 5 mrad (FWHM) was obtained using a similar procedure.
  
   \subsubsection{The cross-talk effect}
   A phenomenon present in the old FPD that affected the horizontal position measurements of the 
   ion track is the 
   cross-talk \cite{CARBONE2012}. In fact, the electron avalanche produced by a given
   proportional wire can induce a charge also in pads corresponding to a neighboring wire. 
   Therefore, the resulted charge distributions measured for a given wire may be distorted 
   affecting the final determination of the X$_i$ coordinate.\\
   In order to mitigate this problem some modification in the new FPD design was required.
   The geometry of the wires and strips was changed making the average distance   
   between two adjacent strips
   larger than the distance between each wire and the corresponding strip. 
   The segmented strips were separated from each other by a 2 mm spacer as shown in Fig. \ref{fig:PADview}.

   \begin{figure}
   \includegraphics[width=0.9\textwidth]{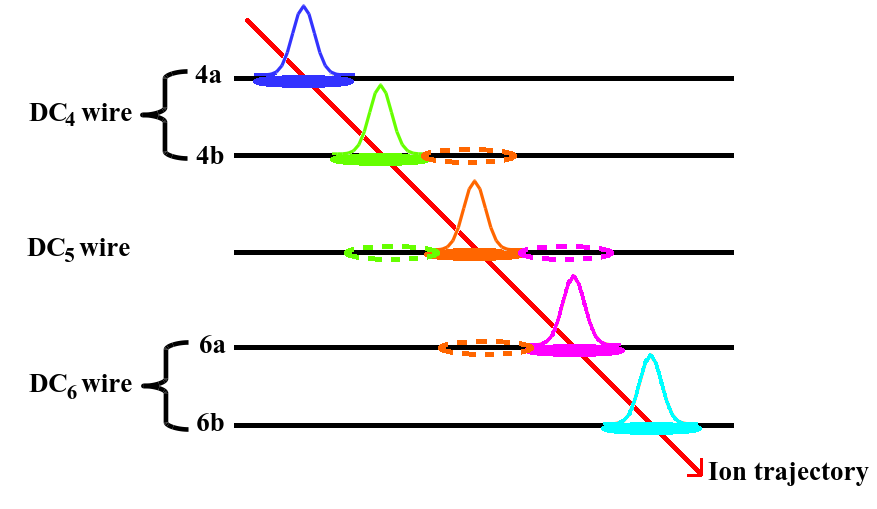}
     \caption{Schematic representation of the induction signal formation in the regions of DC$_4$, 
     DC$_5$ and DC$_6$ multiplication wires. The areas of the main charge induction are denoted with 
     the filled contours, while the areas denoted with the dotted contours correspond to the 
     cross-talk induced signal.}
     \label{fig:inductionformation}
   \end{figure}
\begin{figure}
   \includegraphics[width=0.9\textwidth]{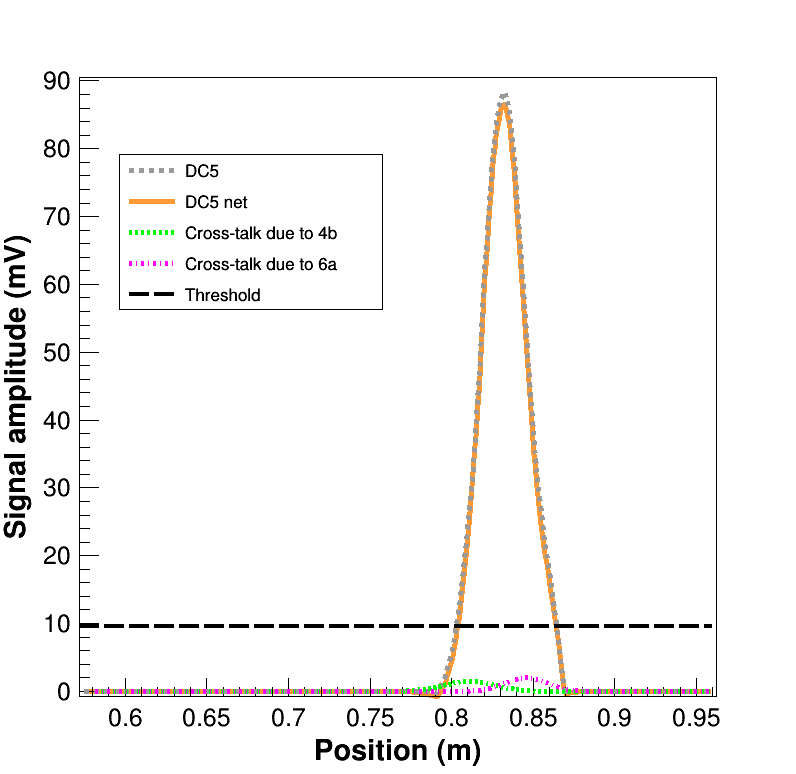}
     \caption{Charge distribution analysis for the induction pads over the DC$_5$ proportional wire. 
     The dashed gray line corresponds to the total charge distribution collected by the induction 
     pads over the DC$_5$ wire. The dashed green and the dashed-dotted magenta lines represent the 
     cross-talk induced signal from DC$_{4b}$ and DC$_{6a}$ wires respectively. The solid orange 
     line corresponds to the signal after subtracting the cross-talk contribution and the dashed 
     black line represents the value of the threshold as it was determined using the COG algorithm
     \cite{CARBONE2012}.}
     \label{fig:dc5_signal}
   \end{figure}

 
   We briefly describe the adopted procedure for estimating 
   the cross-talk induced signal on DC$_5$ strip for a typical event, the same  procedure
   can be extended for the other DCs. As a first step, the measured
   induced charge distributions of DC$_4$ and DC$_6$ wires were fitted each by a Gaussian function. 
   Since DC$_4$ and DC$_6$ are composed by two wires each (see Fig. \ref{fig:PADview}) the charge
   distribution was decomposed  into two internal Gaussian curves, each one describing 
   the induced charge of each wire.
   Then, having determined the height A, the centroid (mean value) and the standard deviation 
   $\sigma$ of the distributions, each one was further decomposed into two Gaussian functions with:
   \begin{equation}
     A_{a}=A_{b}\approx A/2
      \label{eq:cross1} 
   \end{equation}

   \begin{equation}
     \sigma_{a}=\sigma_{b}\approx \sigma
      \label{eq:cross2} 
   \end{equation}
   where, A$_a$, A$_b$ are the heights and $\sigma_a$, $\sigma_b$ are the standard deviations of
   the two Gaussian functions. These so called "internal" Gaussian functions are associated to the 
   signal originating the induced  charge of each  of the two wires composing  DC$_4$
   and DC$_6$, namely, DC$_{4a}$, DC$_{4b}$, DC$_{6a}$ and DC$_{6b}$. In the last step of the 
   analysis, the DC$_{4b}$ and DC$_{6a}$ distributions were superimposed to the measured induced 
   signal of DC$_5$ and were renormalized such as to describe the shape of the distribution.
   The results of the charge distribution analysis for 
   DC$_5$ for a single elastic scattering event with $\theta_{f}$ $\approx$ 62$^o$ are presented 
   in Fig. \ref{fig:dc5_signal}. It is evident that the distortion in the shape of the charge 
   distribution caused by the cross-talk is small. In addition, the amplitude of cross-talk signal 
   is well-below the level of the threshold, which was determined through an iterative procedure by 
   means of the optimized Center Of Gravity (COG) algorithm \cite{CARBONE2012}. Therefore, we may 
   conclude that the cross-talk phenomena are under control.\par
   
   \subsection{Vertical position and angle measurements}

\begin{figure}
      \includegraphics[width=0.9\textwidth]{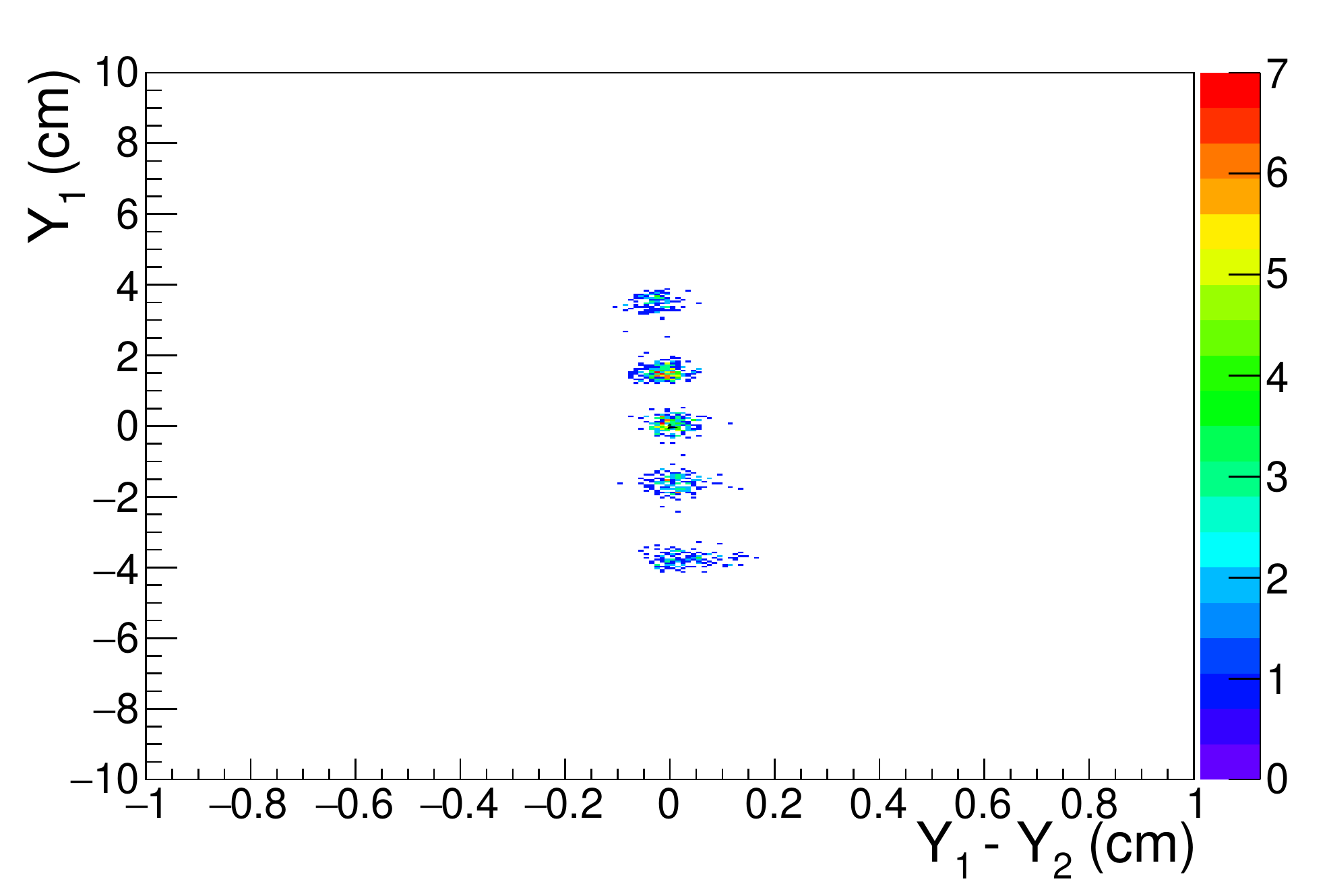}   %
     \caption{y-coordinate versus the Y$_1$-Y$_2$.}
     \label{fig:yversusdy}       %
   \end{figure}

   The Y resolution has been calculated plotting Y$_3$ versus the difference Y$_1$ -Y$_2$ shown in 
   Fig. \ref{fig:yversusdy}. The five spots correspond to the particle passing through each of the 
   five vertical rows of holes of the multi-hole collimator. The FWHM of each spot on the 
   abscissa-projection is 0.6 mm giving a resolution of about 0.4 mm for the single CD.

   The vertical angle $\phi$ is obtained by a linear fit  of the six vertical positions Y$_1$, Y$_2$, 
   Y$_3$, Y$_4$ ,Y$_5$ ,Y$_6$ and the six coordinates Z$_1$, Z$_2$, Z$_3$, Z$_4$ ,Z$_5$ ,Z$_6$ 
   defined by the position of the six drift chambers. That is 
   \begin{equation}
     Z_i = a_0 + a_1 Y_i
   \end{equation}
   where $a_1$ is $tan(\phi)$. For the estimate of the angular resolution we used the formula for 
   the error of the parameters of a linear fit:
   \begin{equation}
     \sigma^2_{a_1} =\Sigma_{i=1}^N \left( \frac{\partial a_1}{\partial y_1}\sigma_y \right)^2 = \sigma^2_y \frac{N}{N\left( \Sigma^N_{i=1}y^2_i \right)- \left(\Sigma^N_{i=1}y_i\right)^2}
   \end{equation}
   where in place of $N \left( \Sigma^N_{i=1}y^2_i \right)- \left(\Sigma^N_{i=1}y_i\right)^2$ the Y
   resolution have been used. The resolution $\delta\phi$ depends on the horizontal angle of the 
   particle trajectory $\theta_f$ since this influences the actual trajectory length inside the 
   tracker. It ranges from 0.3 to 0.7 mrad for trajectories with 
   $\theta_f$=40$^{\circ}$ and $\theta_f$=70$^{\circ}$ respectively.
   We conclude that, also in the vertical direction the position resolution as well the angular
   resolution are better than those of the previous detector.
   
\section{Summary and Conclusions}

The new focal plane detector of the large-acceptance MAGNEX spectrometer has been described 
underlining the innovative aspects relative to the previous FPD. 

It keeps many characteristics of the old FPD, that is the capability to work at pressure ranging 
from few mbar to several tenth of mbar using a  thin entrance window. This guarantees a 
very low detection threshold and the capability to identify particles in a broad range of ionizing condition.
The main characteristics of the new MAGNEX focal plane detector and position and angular resolution 
obtained by using the scattering of $^{18}$O beam of 84 MeV are listed in Tab. \ref{tab:summary}.

The new design based on the segmentation of the gas tracker in six drift chambers all of 
similar size guarantees better performances in terms of tracking precision. 
The track is now sampled in six positions to be compared to the four of the 
old FPD, with a consequent better position and angular resolution of the ion tracks. 
The energy loss the performances have been improved since 
a longer portion of the track in the ionizing gas is sampled compared to
the previous FPD, guaranteeing an higher resolution and therefore a better identification 
capability.
In the design of the new detector, a special care has been given to reduce the effects
of the cross-talk between neighboring strips. In the new FPD the cross-talk has been minimized and
its effects on the track reconstruction are now negligible.

\begin{center}
\begin{table}
  \caption{Main characteristics of the new FPD of MAGNEX spectrometer.}
 \begin{tabular}{l l} 
 \hline\hline
 Energy loss resolution for $^{18}$O&  5\% \\ 
 \hline
 Horizontal position resolution & 0.6 mm \\
 \hline
 Horizontal angular resolution & 2.2 mrad  \\
 \hline
 Vertical position resolution & 0.4 mm \\
 \hline
 Vertical angular resolution & 0.5 mrad  \\
 \hline

 \end{tabular}
 \end{table}
 \label{tab:summary}
\end{center}

\section*{Acknowledgments}
  This project has received funding from the European Union’s
  Horizon 2020 research and innovation program under the
  ERC grant~agreement~GA~714625.

\bibliographystyle{elsarticle-num} 
\bibliography{biblioFPD}

\end{document}